\documentclass[sigconf,authorversion, language=english]{acmart}

\usepackage{array}
\usepackage{adjustbox} 
\usepackage{graphicx}
\usepackage{makecell}
\usepackage{colortbl}

\acmConference[TREC '24]{the 33th Text Retrieval Conference}{Nov 18--22,
  2024}{Rockville, MD}
\acmISBN{978-1-4503-XXXX-X/18/06}

\begin{document}


\title{RALI@TREC iKAT 2024: Achieving Personalization via Retrieval Fusion in Conversational Search}


\author{Yuchen Hui}
\orcid{0000-0002-9659-3714}
\affiliation{%
  \institution{RALI, Université de Montréal}
  \city{Montréal}
  \state{Québec}
  \country{Canada}
}
\email{yuchen.hui@umontreal.ca}

\author{Fengran Mo}
\orcid{0000-0002-0838-6994}
\affiliation{%
  \institution{RALI, Université de Montréal}
  \city{Montréal}
  \state{Québec}
  \country{Canada}
}
\email{fengran.mo@umontreal.ca}

\author{Milan Mao}
\affiliation{%
  \institution{RALI, Université de Montréal}
  \city{Montréal}
  \state{Québec}
  \country{Canada}
}
\email{milan.mao@umontreal.ca}

\author{Jian-Yun Nie}
\orcid{0000-0003-1556-3335}
\affiliation{%
  \institution{RALI, Université de Montréal}
  \city{Montréal}
  \state{Québec}
  \country{Canada}
}
\email{nie@iro.umontreal.ca}

\renewcommand{\shortauthors}{Hui, et al.}
\begin{abstract}
The \textbf{R}echerche \textbf{A}ppliquée en \textbf{L}inguistique \textbf{I}nformatique (\textbf{RALI}) team participated in the 2024 TREC Interactive Knowledge Assistance (iKAT) Track. In personalized conversational search, effectively capturing a user's complex search intent requires incorporating both contextual information and key elements from the user profile into query reformulation. The user profile often contains many relevant pieces, and each could potentially complement the user's information needs. It is difficult to disregard any of them, whereas introducing an excessive number of these pieces risks drifting from the original query and hinders search performance
This is a challenge we denote as over-personalization. To address this, we propose different strategies by fusing ranking lists generated from the queries with different levels of personalization.

\end{abstract}




\maketitle

\section{Introduction}
Personalized conversational information retrieval (CIR) systems aim to deliver tailored search results for users' specific information needs through multi-turn dialog, requiring the model to consider the previous user-system interactions and user profiles. The key challenge of personalized CIR lies in simultaneously extracting relevant information from the contextual history and the user profile, and then combining these elements to conduct effective retrieval via understanding the user's complex search intent.

In CIR, the user's contextual search intent can be distilled into a stand-alone query through conversational query reformulation techniques \cite{canard}. 
In personalized CIR~\cite{aliannejadi2024trec}, it is intuitive to reformulate the query to be both context and persona independent, allowing better encapsulation of the user's nuanced search intent.
A recent study \cite{mo2024leverage} has demonstrated that a de-contextualized and personalized query can be built by leveraging language models (LLMs) as a search intent interpreter. However, while LLMs have shown remarkable success in conversational query reformulation (CQR) \cite{llm4cs}, the stand-alone queries produced by selecting and incorporating relevant pieces in the user profile are not always good search queries. This is because user profile terms are inherently noisier than those extracted from conversation history. Terms added by LLMs in CQR typically address missing pieces caused by coreference and ellipsis—key components that contribute to the majority of the query's semantic meaning. They are crucial to make the user's primary search intent understandable. On the other hand, terms from user profiles are usually only weakly semantically related to the query. Although these profile terms may hint at the user's preferences and potentially complement the search intent, explicitly expanding them into the reformulated queries might lead to the query drift issue, i.e., retrieving irrelevant documents that focus solely on these terms. This creates a dilemma: excluding profile terms risks omitting valuable personalized context, whereas introducing an excessive number of these pieces risks drifting from the original query and hinders search performance. This is a challenge we denote as over-personalization, which could be illustrated by the following de-contextualized and personalized query example:
\textit{What Turkish souvenir do you suggest for my mom, considering she has a collection of antique crystals and porcelains?}
where ``\textit{collection of antique crystals and porcelains}'' originated from the user's profile. Although the reformulated query condition on user profiles indeed enriches the user's search intent by implying a preference for fine art pieces as souvenirs, it also introduces potential noise. Highly relevant documents addressing both ``\textit{Turkish souvenir}'' and ``\textit{crystals and porcelains}'' should be ranked at the top. However, many irrelevant documents that focus solely on ``\textit{crystals}'' and ``\textit{porcelains}'' without mentioning ``\textit{Turkish souvenir}'' may also rank high. To address this issue, one may consider distinguishing highly relevant documents from irrelevant ones among the top-ranked results. To achieve this, note that these highly relevant documents would also score well when retrieved using a de-contextualized but non-personalized query. This suggests that combining document scores from both personalized and non-personalized queries can help identify truly valuable candidates: all those performing well in both ranking lists. 

This simple yet practical principle is named ``\textbf{\textit{The Chorus Effect}}'' \cite{vogt1999fusion}. Introduced in retrieval fusion approaches, it states that when several retrieval methods suggest that an item is relevant to a query, this tends to be stronger evidence for relevance. Dating back to the 20th century, retrieval fusion methods have already exploited this principle by merging search results from different retrieval systems to improve performance.

In this year's iKAT track, we investigate the effectiveness of retrieval fusion applied to personalized and non-personalized context-independent queries rewritten by LLMs.

\section{Related Works}
\textbf{Conversational Search.} Conversational search is an information-seeking process through interactions with a conversation system \cite{cir_book,mo2024survey}. Existing methods can be roughly categorized into two groups: conversational query rewriting (CQR) and conversational dense retrieval (CDR). Conversational dense retrieval \cite{chatretriever,or_convqa,mao2023learning,mo2024aligning,mo2024convsdg} directly encodes the whole conversational search session to perform end-to-end dense retrieval. On the other hand, CQR methods aim to produce a stand-alone de-contextualized query that can be then submitted to any ad-hoc search models for retrieval purpose. Existing studies try to select useful tokens from the conversation
context \cite{fang-etal-2022-open, kumar2020making, voskarides2020query,mo2023learning} or train a generative rewriter model with conversational sessions to mimic the human rewrites \cite{lin2020conversational, vakulenko2021question, yu2020few}. To optimize query rewriting for the search task, some studies adopt reinforcement learning \cite{chen-etal-2022-reinforced, Wu2021CONQRRCQ} or apply the ranking signals with
the rewriting model training \cite{mao2023search, mo-etal-2023-convgqr}, while others jointly learn query rewriting and context modeling \cite{qian2022explicit}. Rermarkably, some recent methods are proposed to directly prompt the LLMs to generate rewrites \cite{llm4cs, mo2024leverage, mo-etal-2024-chiq, ye-etal-2023-enhancing}. In our submissions, contextual and personalized elements are incorporated into current turn user queries by LLM-based CQR methods.

\noindent \textbf{Retrieval fusion.} Retrieval fusion utilizes a group of retrieval strategies (e.g., different information retrieval systems or query variations) to search in the same document collection, then merges the resulting ranking lists to improve retrieval effectiveness. 
Retrieval fusion methods are typically categorized into rank-based methods and score-based methods. Rank-based methods, such as RRF \cite{rrf}, Round-Robin fusion \cite{voorhees1995collection}, Borda count \cite{borda_fuse}, and Condorcet fusion \cite{montague2002condorcet}, rely solely on the order of documents in multiple ranking lists. Score-based methods, by contrast, require relevance scores from candidate lists. Among these, the linear combination method (LC) is highly flexible, allowing different weights for each retrieval strategy. In LC, the relevance score of a document in the fused list is calculated as a weighted average of its scores in all candidate lists, where it is crucial to determine an appropriate weight assignment. Methods like CombSum and CombMNZ \cite{fox1994combination} assign equal weights to all candidate lists, while more sophisticated weighting assignment usually relies on relevance judgment. \citet{vogt1999fusion} and \citet{wu2009assigning} assign weights according to the performance of candidate retrieval strategies on a group of training query. i.e., the weight of a strategy is based on its average performance on training queries using specific metrics like Mean Average Precision. \citet{wu2012linear} further employs multiple linear regression, using binary relevance judgments as the dependent variable and candidate list scores as input variables, with regression coefficients serving as weights. A comprehensive review of fusion in ad-hoc search is available in \cite{kurland2018fusion}. Beyond ad-hoc search, retrieval fusion has other application scenarios in the broad context of IR. For example, it is employed by \cite{gialampoukidis2016hybrid} for combining multi-modal information in multimedia retrieval. \cite{bruch2023analysis, wang2021bert} explore fusion strategies to combine semantic match by dense retrievers \cite{karpukhin-etal-2020-dense} with lexical match by sparse retrievers. 
A recent attempt at retrieval fusion in personalization is that \citet{huang2024unleashing} trains a personalized Reinforcement Learning policy for weight assignment in the scenario of linearly fusing candidate item lists in the multi-channel recall phase of a recommender system. 
In personalized conversational search, it is worth exploring other reasonable weight assignment approaches since no public training data is available~\cite{mo2024leverage}. 
\section{Methodology}
\subsection{Manual runs}
We submitted two manual runs, both using the two-stage retrieval-reranking paradigm. In both stages, only the provided human rewrites are used as the search query for each conversational turn. BM25 serves as the retriever for both manual runs. In the reranking stage, we compare the performance of the MonoT5\footnote{Checkpoint available at castorini/monot5-base-msmarco-10k} \cite{monot5} and RankLlama\footnote{Checkpoint available at castorini/rankllama-v1-7b-lora-passage } \cite{rankllama} rerankers in the out-of-domain evaluation setting of the iKAT track. Top 50 documents recalled by BM25 are reranked. We did not participate in the response generation task for these manual runs.
\subsection{Automatic runs}
Our automatic runs aim to evaluate the effectiveness of a personalized ranking list fusion approach. We submitted four automatic runs in total, though our analysis here focuses on two key runs, as the others do not incorporate PTKB. Details are as follows:
\subsubsection{\textbf{RALI\_gpt4o\_fusion\_rerank}}
This run involves four main steps:
\begin{enumerate}
    \item \textbf{Query Reformulation}. For each turn, three types of query reformulations are generated via prompting the GPT-4o\footnote{The version is gpt-4o-2024-08-06} model:\begin{itemize}
        \item A de-contexualized but non-personalized query rewrite;
        \item The above query concatenated with a GPT-4o-generated response as a form of expansion;
        \item A de-contextualized and personalized query rewrite.
    \end{itemize}
    To obtain the first query and the response used in the second query, we apply the \textit{Rewriting-And-Response} prompt from the LLM4CS framework \cite{llm4cs}. The third query is generated using a Chain-of-Thought prompt guiding GPT-4o to incorporate relevant PTKB elements. This prompt includes human-annotated reasoning processes that analyze how PTKB elements are integrated into human rewrites in iKAT-23 test conversations.
    \item \textbf{Ranking List Fusion.} This component is central to our approach. For each conversational turn, we first get three ranking lists by querying BM25 with the reformulated queries from the previous step. Each ranking list contains the top 1000 documents retrieved for that turn. A score-based linear combination method is then used to fuse these lists into a single ranking list. Specifically, for a document $D$, its score in the final ranking list is computed as:
    $$
    S_{final}(D) = \alpha_1 S_{1}(D) + \alpha_2 S_{2}(D) + \alpha_3 S_{3}(D)
    $$
    Where $S_n(D)$ is the score for document D in ranking list $n$ and $ \alpha_n$ is the fixed weight assigned to each type of query reformulation. Note that the weights $\alpha$ are set consistently across all conversational turns for each reformulation type. If a document $D$ does not present within the top 1000 documents in ranking list n, $S_n(D)$ defaults to the score of the 1000th document in that list. The weights are optimized based on performance tuning on the iKAT-23 test collection.
    \item \textbf{Reranking}. The top 50 documents from the fused ranking list are reranked using the MonoT5\footnote{Checkpoint available at castorini/monot5-base-msmarco-10k} \cite{monot5} reranker. Here, the de-contextualized and personalized query generated in the first step is regarded as the search query for reranking.
    \item \textbf{Response Generation}. The top three documents from the reranked list, along with the conversational context and the user's PTKB, are provided to GPT-4o to generate a response.
    
\end{enumerate}

\subsubsection{\textbf{RALI\_gpt4o\_fusion\_norerank}}
This run uses the same query reformulation and BM25 ranking list fusion methods as described in the previous run. However, it does not rerank the fused ranking list or generate responses.

\section{Evaluation Results}
\begin{table*}[t]
    \centering
    \caption{TREC iKAT 2024 evaluation results}
    \vspace{-2ex}
    \begin{tabular}{lcccccccc}
    \toprule
        Submissions & MRR & NDCG@5 & NDCG@10 &  Recall@10  & Recall@100 & MAP@1K   \\ 
        \midrule
        \rowcolor{gray!20} \multicolumn{7}{c}{automatic runs} \\
        BM25 fusion + no rerank         & 75.2 & 37.9 & 35.6 & 7.37 & \textbf{28.4} & 18.1\\
        BM25 fusion + monoT5            & \textbf{84.1} & \textbf{51.1} & \textbf{47.7} & \textbf{9.72} & \textbf{28.4} & \textbf{21.1} \\
        \rowcolor{gray!20} \multicolumn{7}{c}{manual runs} \\
        BM25 + monoT5         & 79.8 & 40.4 & 35.4 & 7.7 & 20.2 & 13.6\\ 
        BM25 + rankllama      & 80.0 & 42.3 & 36.7 & 7.7 & 20.2 & 13.6\\    
        \bottomrule
     \end{tabular}
     \label{table: main_results}
\vspace{-2ex}
\end{table*}
The evaluation results for passage ranking task of the afore-mentioned four submissions are presented in Table \ref{table: main_results}. Among the four methods, the run combining BM25 ranking list fusion and monoT5 reranking achieves the best performance, proving the effectiveness of retrieval-fusion. 
The evaluation results for response generation task has not been made available yet.

It is worth noting that we also evaluated all our submission methods on the iKAT-23 test collection, and the results show that the automatic fusion-then-reranking approach could not outperform runs based on manual human rewrites. Another similar interesting finding is that RankLlama outperforms MonoT5 this year, whereas the opposite is true on last year’s test collection. A possible explanation for this discrepancy between the two years is \textbf{assessment bias}. Specifically, iKAT test collections are built using pooling, which tends to advantage participant methods since more documents retrieved by them are assessed. Last year, neither RankLlama-based reranking nor our proposed data fusion approach participated, which may explain why they can not achieve comparable performance to this year. See more detailed discussions in Section \ref{assessment_bias}.

\section{Assessment Bias}
\label{assessment_bias}
During our exploration for better personalization strategies, various types of query reformulation are tested for BM25 lexical search. Surprisingly, according to our manual observation, although some of these reformulated queries express the search intent better than human rewrites, they do not perform as well as the latter. 
This counter-intuitive finding drives us to investigate further via case studies. Specifically, we expanded human rewrites manually by adding clearly relevant terms and then compared the performance of these expanded versions with the original rewrites. Intuitively, the improved queries is expected to outperform the originals; However, the original ones always produce superior search results.
\begin{table}[H]
    \centering
    \caption{Original and expanded human rewrite for iKAT 2023 test turn 9-1-3 utterance }
    \vspace{-2ex}
    \resizebox{\columnwidth}{!}{ 
    \begin{tabular}{lcccccccc}
    \toprule
        Variation &  Query Reformulation\\ 
        \midrule
        Original & What about the DASH diet? I heard it is a healthy diet.\\
        Expanded & \makecell{What about the DASH diet? I heard it is a healthy diet. \\ + \textbf{Dietary Approaches to Stop Hypertension}} \\
    \bottomrule
    \end{tabular}
    }
    \label{table: case study}
\vspace{-2ex}
\end{table}

An example is shown in Table \ref{table: case study}. The original human rewrite for utterance 9-1-3 from iKAT 2023 asks about a healthy diet named ``\textit{DASH}'', which stands for ``\textit{Dietary Approaches to Stop Hypertension}''. Generally, concatenating a query with the full name of an acronym in it would help BM25 to retrieve more relevant documents. However, evaluation results in Table \ref{table: case study eval} indicate that these conceptually reasonable expansions drastically harm retrieval metrics calculated based on the relevance judgment.

\begin{table}[H]
    \centering
    \caption{Evaluation and statistics on the original and expanded human rewrite}
    \vspace{-2ex}
    \resizebox{\columnwidth}{!}{ 
    \begin{tabular}{lcccccccc}
    \toprule
        Variation &  NDCG@10 & R@20 & \# Assessed docs in Top20\\ 
        \midrule
        Original & 52.9 & 0.12 & 14\\
        Expanded & 29.4 & 0.04 & 5 \\ 
    \bottomrule
    \end{tabular}
    }
    \label{table: case study eval}
\vspace{-2ex}
\end{table}
We attribute the unexpectedly poor performance of improved query reformulation to biases in the iKAT track's assessment process. The iKAT track has two main submission classes, \textit{automatic} and \textit{manual}. All manual runs utilize the human rewrite as an input query, hence more documents retrieved by human rewrites are assessed in the TREC pooling process. This fact places all other query reformulations at a disadvantage as the assessment is less complete for them. For instance, in the turn 9-1-3 case, only 5 of the top 20 documents recalled by our more reasonable expanded query are assessed, compared to 14 for the original human rewrites. This bias might explain the inferior performance of the expanded 9-1-3 human rewrite.

A similar observation is reported in \cite{monot5}, where the authors find that due to the use of early year TREC test collections built before the advent of BERT-based retrieval models, documents retrieved by their T5-based models have lower proportions of judgment rate compared to those searched by BM25.

These raise questions about the reusability of iKAT test collections: can the pooling-built iKAT test collections reliably evaluate non-participant retrieval strategies that did not contribute to the relevance judgments? Investigations conducted by \cite{voorhees2022can} confirmed that a test collection is reusable for new retrieval methods if \textbf{\textit{deep pools}} are used over highly effective runs during assessment. This conclusion is supported by experiments on the TREC-8 ad hoc collection, which are created with a large pooling depth of 100, contrasting with the smaller pooling depth of 10 for iKAT 2023, as shown in Table \ref{table: test_collection_statistics}. In the scenario of shallow pooling however, a separate study \cite{ test_collection_bias} on TREC 2019 Deep Learning track \cite{craswell2020overview} shows that, with depth of 10, the reliability of evaluation results using such test collections can not be guaranteed, aligning with our observations for iKAT 23. 
\begin{table}[H]
    \centering
    \caption{Statistics of test collections}
    \vspace{-2ex}
    \resizebox{\columnwidth}{!}{ 
    \begin{tabular}{lcccccccc}
    \toprule
        Collection &  \# Query & Pool Depth & \# Assessed docs/psgs & Collection size \\ 
        \midrule
        TREC-8 & 50 & 100 & 86,830 & 525,000 \\
        TREC DL-19 & 43 & 10 & 9,260 & 8,841,823 \\
        iKAT-23 & 174 & 10 & 26,159  & 116,838,987 \\ 
        iKAT-24 & 116 & up to 30 & 20,575 & 116,838,987 \\ 
    \bottomrule
    \end{tabular}
    }
    \label{table: test_collection_statistics}
\vspace{-2ex}
\end{table}
Further more, \cite{buckley2007bias} found that a biased relevance set may occur as the size of the document collection increases. Notably, as shown in Table \ref{table: test_collection_statistics}, the iKAT test collections has a tremendous document set while remaining a low pooling depth --- precisely the case identified in that study as susceptible to assessment bias.

Deeper pooling depth may help alleviate the assessment bias problem, but it is a costly remedy. Fortunately, regardless of a biased test collection, if a study claims that a new method outperforms TREC participant results, this conclusion remains valid despite the bias. Any unfairness in the assessment would only reduce the metrics for the new method rather than inflate them. Nevertheless, it remains valuable to explore innovative pooling and evaluation practices that could lead to more reusable test collections~\cite{soboroff2024don}.

\section{Conclusion}
In this paper, we present our solution for the 2024 TREC iKAT passage ranking task, with a focus on achieving personalization through retrieval fusion in conversational search. The overall evaluation results demonstrate the effectiveness of retrieval fusion. 
Additionally, we discuss the assessment bias and reusability of the iKAT test collections. Future work will involve exploring more effective fusion strategies.


\bibliographystyle{ACM-Reference-Format}
\bibliography{sample-base}



\end{document}